\documentclass[12pt]{iopart}

\usepackage{ dsfont }
\usepackage{hyperref}
\usepackage{graphicx}
\newcommand{\ddt}[1]{\frac{d #1}{dt}}
\newcommand{\hmone}[1]{\|\nabla^{-1} #1\|_{L^{2}}}
\newcommand{\ltwo}[1]{\|#1\|_{L^{2}}}
\newcommand{\hone}[1]{\| \nabla #1\|_{L^{2}}}

\newcommand{\sint}[1]{\int_{D} #1 \, d^{d}\mathbf{x}}

\renewcommand{\vec}[1]{\mathbf{#1}}
\newcommand{\linf}[1]{\| #1 \|_{L^{\infty}}}

\renewcommand{\u}{\mathbf{u}}
\newcommand{\ppt}[1]{\partial_{t} #1}
\newcommand{\lap}{\Delta }
\newcommand{\invlap}{\Delta^{-1}}
\newcommand{\pbrac}[1]{\left( #1 \right)}
\newcommand{\sbrac}[1]{\left[ #1 \right]}

\begin{document}

\title[Diffusion-limited mixing by incompressible flows]{Diffusion-limited mixing by incompressible flows}

\author{Christopher J. Miles$^{1,2,3}$ and Charles R. Doering$^{1,2,3}$ }

\address{$^1$ Department of Physics, University of Michigan,
Ann Arbor, MI 48104-1040, USA}
\address{$^2$ Department of Mathematics, University of Michigan,
Ann Arbor, MI 48104-1043, USA}
\address{$^3$ Center for the Study of Complex Systems, University of Michigan,
Ann Arbor, MI 48104-1107, USA}
\ead{doering@umich.edu}
\vspace{10pt}
\begin{indented}
\item[]\today
\end{indented}

\begin{abstract}
Incompressible flows can be effective mixers by appropriately advecting a passive tracer to produce small filamentation length scales. In addition, diffusion is generally perceived as beneficial to mixing due to its ability to homogenise a passive tracer. However we provided numerical evidence that, in the case where advection and diffusion are both actively present, diffusion produces nearly neutral or even {\it negative} effects by limiting the mixing effectiveness of incompressible optimal flows. This limitation appears to be due to the presence of a limiting length scale given by a generalised Batchelor length \cite{Batchelor1959a}. This length scale limitation in turn affects long-term mixing rates. More specifically, we consider local-in-time flow optimisation under energy and enstrophy flow constraints with the objective of maximising mixing rate performance. We observe that, for enstrophy-bounded optimal flows, the strength of diffusion has no impact on the long-term mixing rate performance. For energy-constrained optimal flows, however an increase in the strength of diffusion decreases the mixing rate. We provide analytical lower bounds on mixing rates and length scales achievable under related constraints (point-wise bounded speed and rate-of-strain) by extending the work of Z. Lin {\it et al.} \cite{JFM2011} and C.-C. Poon \cite{Chi-Cheu1996}. 
\end{abstract}
%
\vspace{2pc}

%
%
%

\section{Introduction}
\label{sec:introduction}

Mixing of a passive tracer quantity, such as temperature, solute concentration, or salinity, by an incompressible flow is a fundamental fluid process. It is relevant to many domains such as turbulence theory \cite{Dimotakis2005,Violeau2000a}, aerospace engineering, oceanography \cite{Wunsch2004}, and atmospheric sciences. It also serves as a key industrial process within the food, pharmaceutical, petrochemical, and other industries \cite{paul2004handbook}. Although mixing is highly prevalent and used, its fundamental principles are still not fully known.

The effect of diffusion and advection on the rate of fluid mixing depends on the unique fluid properties, specific mixing flow, and boundary geometry. Thus general principles of mixing applicable to the vast variety of mixing situations would be beneficial. In particular, it is valuable to determine how the mixing rate (typically the most optimal mixing rate) depends on aggregate flow intensity measures such as energy and enstrophy. This is the objective of the research program encompassing many efforts \cite{CS2013,GI2014,JLT2012,JFM2011, Miles2017a,  JLT2012, DF2014, GM2005} in last decade. 

With these goals in mind, a common approach taken throughout the literature is to consider the evolution of a tracer quantity $\theta (\vec{x},t)$ advected by an incompressible ($\nabla \cdot\vec{u}=0$) flow $\vec{u}$ with mild physical constraints within a periodic box  of side length $L$ in $d$ dimensions. The tracer concentration field $\theta$ evolves according to the advection-diffusion equation,
\begin{equation}
	\label{eq:PDE_advection}
	\ppt{\theta}+\mathbf{u}\cdot \nabla \theta=\kappa \lap\theta,
\end{equation}
with initial data $\theta(\mathbf{x},0)=\theta_{0}(\mathbf{x})$, where $\kappa$ is the molecular diffusion coefficient. The flow is constrained by enstrophy $\ltwo{\nabla\u} = \Gamma L^{d/2}$ or energy $\ltwo{\u} = UL^{d/2}$ where $\Gamma$ is the root mean square rate-of-strain and $U$ is the root mean square speed. 

\begin{figure}
	\centering
	\includegraphics[width=0.5\textwidth]{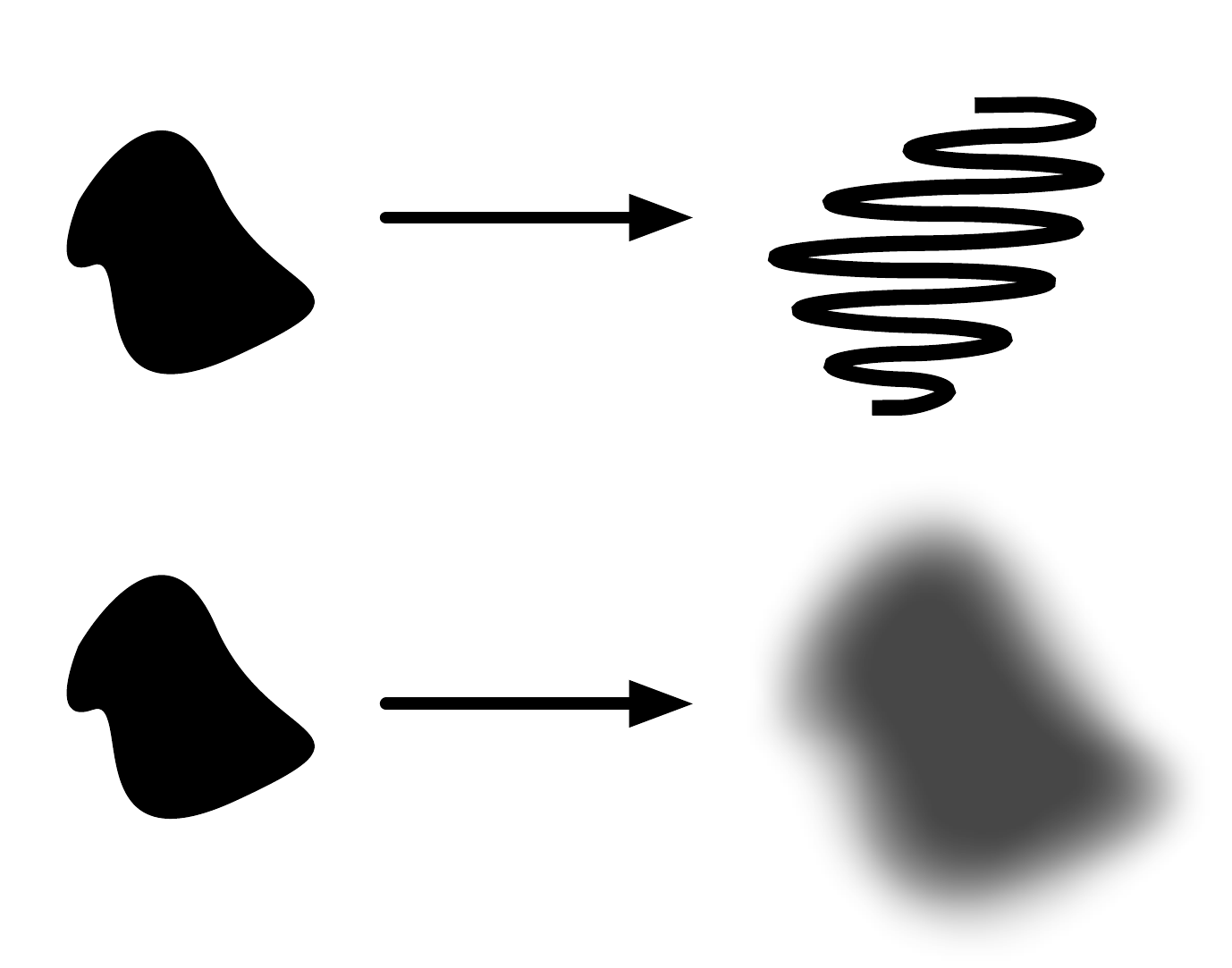}
	\caption{Filamentation is the reduction of the scale of segregation as illustrated by the top transformation. Homogenisation is the reduction of the intensity of segregation as illustrated by the bottom transformation.}
	\label{fig:scale-and-intensity}
\end{figure}

We also need a measure of `mixed-ness'. In the pioneering work of P. V. Danckwerts (1952)\cite{Danckwerts1952}, the author identifies two measures of mixedness: the scale of segregation and the intensity of segregation. The {\it scale of segregation } quantifies the characteristic length scale present in the tracer concentration $\theta$. For instance, the process of thinning, elongating, and folding of a blob, as seen in the top graphic of figure \ref{fig:scale-and-intensity}, reduces the scale of segregation by creating a rich maze-like pattern with thin strands of dye. We will refer to mixing by the reduction of the scale of segregation as {\it filamentation}.  On the other hand, the {\it intensity of segregation } quantifies the variation of the concentration amplitude. We will refer to mixing by the reduction of the intensity of segregation as {\it homogenisation}. This process is illustrated by the bottom graphic of figure \ref{fig:scale-and-intensity}.

 $H^{-1}$ norm or mix-norm \cite{GM2005} is a single measure of mixing that accounts for both the scale and intensity of segregation. This is a common measure of mixing throughout the literature and is defined by  
\begin{equation}
\hmone{\theta}=\sqrt{\sint{ |\nabla^{-1} \theta( \vec{x},t)|^2}}=\sqrt{ \sum_{\vec{k}\neq \vec{0}} L^d \frac{|\hat{\theta}_{\vec{k}}(t)|^{2}}{|\vec{k}|^2}}
\end{equation}
where $\nabla^{-1}=\nabla \Delta^{-1}$, the operator $\Delta^{-1}$ acting on  a function $\rho$ returns the solution $\phi$ of the Poisson equation $ \Delta \phi = \rho $, $\hat{\theta}_{\vec{k}}(t) =  \frac{1}{L^{d}}\sint{\theta(\vec{x},t)e^{-i\vec{k}\cdot\vec{x}}}$, and $\vec{k}$ is the wave number.  Lower values of the  $H^{-1}$ norm correspond to a more mixed state. Note that $H^{-1}$ norm can decrease in two ways.  The first way is by transferring spectral mass from the lower wave numbers to the higher wave numbers to take advantage of the $1/|\vec{k}|^2$ dependence. This produces a scalar field with sharp gradients and small length scales. This corresponds to the reduction of scale of segregation or filamentation. The second way is to decrease the Fourier amplitudes $|\hat{\theta}_{\vec{k}}|$ for $\vec{k}\neq 0$. This corresponds to a reduction in the intensity of segregation, or homogenisation. Thus we can see that the $H^{-1}$ embodies both senses of mixing. 

The $L^{2}$ norm $\ltwo{\theta}$ and the $H^{1}$ norm $\hone{\theta}$ are also common measures of mixing and will be considered here as well.  For those interested in other measures of mixing, see \cite{JLT2012}. 

In the case without advection ($\vec{u}=\vec{0}$), equation \eref{eq:PDE_advection} reduces to the classical heat equation \cite{Evans2010} and the Fourier modes evolve according to $\hat{\theta}_{\vec{k}}(t)=\hat{\theta}_{\vec{k}}(0)e^{-\kappa|\vec{k}|^2t}$. Thus we have explicit analytical results for the decay of the $H^{-1}$ mix-norm by simply substituting this result. Note that $H^{-1}$ mix-norm will decay monotonically. Diffusion is unable to transfer spectral mass from the low wave number modes to the high wave number modes and thus is incapable of filamentation. Thus the pure diffusion case solely exploits homogenisation. Also notice the unequal weighting attach to each mode. The Fourier modes with large wave number $|\vec{k}|$ decay at a faster rate relative to the decay of those with small wave number.


In the case without diffusion ($\kappa = 0$), pure advection of the flow is the only method of mixing, colloquially known as stirring. For a flow that is constrained by enstrophy, the mix-norm decays at most exponentially where the exponential rate is proportional to $\Gamma$ \cite{GI2014,CS2013}. This was mathematically proven by two separate approaches: G. Iyer {\it et al.} \cite{GI2014} used regularisation results \cite{Crippa} of partial differential equations  while C. Seis \cite{CS2013} used methods from optimal transportation theory \cite{villani2003topics}. Furthermore, enstrophy-constrained flows that realise this exponential decay rate have been constructed analytically \cite{Alberti2014a}. On the other hand, energy-constrained flows can achieve even faster mixing rates. In fact they can achieve {\it perfect mixing in finite time} which means that the $H^{-1}$ mix-norm approaches zero in finite time as opposed to approaching zero in infinite time as in the case for enstrophy-constrained mixing. This can be demonstrated by a `chequerboard' flow \cite{JMP2012} where the mix-norm achieves perfect mixing in finite time with a linear decay rate.  In either flow intensity constraint, note that $H^{-1}$ mix-norm decreases by exclusively exploiting filamentation without homogenisation. This is  exactly opposite to the purely diffusive case.

Finally, the case with diffusion and advection is the least explored in this framework and the focus of this paper. D. Foures {\it et al.} \cite{DF2014} showed that the evolution of the $H^{-1}$ and $L^2$ norms decrease monotonically under the chequerboard flow introduced by E. Lunasin {\it et al.} \cite{JMP2012} while the $H^{1}$ increases until it reaches a peak and then decreases. This peak corresponds to a  time when the length scales developed are small enough for diffusion to effectively act on steep gradients. C. Miles and C. Doering \cite{Miles2017a} explored this phenomena further in the context of a shell model, a reduced model using ordinary differential equations that mimic the spectral dynamics of the advection-diffusion equation. The authors concluded that shell-model mixing could not surpass length scales given by $\sqrt{\kappa/ \Gamma}$ for enstrophy-constrained flows and $\kappa/U$ for energy-constrained flows up to $O(1)$ constants. These length scales can be identified as a generalised Batchelor scale \cite{Batchelor1959a}, introduced in the context of turbulence theory. The limitations on the length scale controlled the long-run optimal mixing rates determined to be $\Gamma$ under the enstrophy constraint and $U^2/\kappa$ under the energy constraint up to $O(1)$ constants. In contrast to the `pure' cases mentioned earlier, it is important to note that the $H^{-1}$ can now decrease by the two avenues of homogenisation and filamentation simultaneously. 

At this point, we can already see a glimpse of a conflict between diffusion and advection for the ultimate goal of optimal mixing. Pure advection succeeds at filamentation by transferring spectral mass from the low wave number modes to the high wave number modes in a continuous fashion. However in the presence of diffusion, a once optimal pure advection flow exceptional at filamentation will be met with potential conflict since homogenisation by diffusion may stifle its progress in transferring spectral mass to high wave number modes. Given that diffusion is ubiquitous, we must come to terms with this conflict to produce efficient mixing.

In this paper, we make progress towards answering ``What is the most optimal mixing rate in the presence of diffusion for an enstrophy or energy constrained flow?'' This question was also asked in the context of the shell model. We would like to determine if the predictions of the shell model hold in the partial differential equation setting. 

We approach the posed question by considering the general setup introduced earlier of the evolution of passive scalar in a periodic box. We consider the local-in-time optimisation problem first introduced by Z. Lin {\it et al.} \cite{JFM2011}, but now in the context of diffusion. Local-in-time optimisation seeks to find the optimal flow that achieves the best instantaneous mixing rate. We will see that the best choice leads to a $\vec{u}$ that depends on $\theta$. This feedback causes the dynamics of $\theta$ governed by \eref{eq:PDE_advection} to be nonlinear.

We will demonstrate that homogenisation via diffusion and filamentation via advection can sometimes be in conflict and collectively produce a negative impact on mixing. We show numerical evidence that filamentation length scale appears to be limited by the Batchelor scale as seen in the shell model, even when actively trying to choose the most optimal flow to enhance filamentation. Thus, this may suggest that the Batchelor scale does not only limit turbulent flows but also all incompressible flows under the flow constraints considered here. Although these quantities have been known in the context of turbulence theory, the impact of these limitations on mixing rates has not been fully studied to our knowledge.

The paper is organised as follows. We introduce the necessary theory regarding local-in-time optimisation, a shell model, and $L^{\infty}$ flow constraints in section \ref{sec:theory}. Section \ref{sec:numerical_experiment} details the methodology and results of numerically implementing local-in-time flow optimisation. Lastly, we finish with a discussion and conclusion in sections \ref{sec:discussion} and \ref{sec:conclusion} respectively.


\section{Theory}
\label{sec:theory}
\subsection{Local-in-time flow optimisation}

We will consider the evolution of a tracer quantity $\theta$ governed by equation \ref{eq:PDE_advection} under an the incompressible flow $\vec{u}$.  Recall the flow is constrained by enstrophy $\ltwo{\nabla\u} = \Gamma L^{d/2}$ or energy $\ltwo{\u} = UL^{d/2}$ where $\Gamma$ is the root mean square rate-of-strain and $U$ is the root mean square speed. 

For the enstrophy-bounded flow problem, we choose the same length scale $L$, the velocity scale $L\Gamma $, and  the time scale $1/\Gamma$. For the energy-bounded flow problem, we non-dimensionalise the system by choosing $L$ as the length scale, $U$ as the velocity scale, and $L/U$ as the time scale.  Both scalings produce the following form of the advection-diffusion equation,
\begin{equation}
\label{eq:nd_ade}
	\ppt{\theta}+\mathbf{u}\cdot \nabla \theta=\frac{1}{Pe} \lap\theta,
\end{equation}
where $Pe=  \frac{\Gamma L^2}{\kappa}$ for the enstrophy-constrained case and $Pe= \frac{UL}{\kappa}$ for the energy-constrained case.   

We consider the local-in-time optimisation strategy first introduced by Lin {\it et al.} \cite{JFM2011} in the case without diffusion. We find that this strategy generalises to the case with diffusion. The local-in-time optimal velocity fields maximise the instantaneous mixing rate by minimising $\ddt{}\hmone{\theta}^2$. The optimal velocity fields are given instantaneously for the enstrophy case by (in non-dimensional form)
\begin{equation}
\mathbf{u}= \frac{-\invlap\mathds{P}(\theta \nabla \invlap\theta)}{\langle |\nabla^{-1}\mathds{P}(\theta \nabla \invlap\theta)|^2\rangle^{1/2}}
\end{equation}
and for the energy case by 
\begin{equation}
\mathbf{u}= \frac{\mathds{P}(\theta \nabla \invlap\theta)}{\langle |\mathds{P}(\theta \nabla \invlap\theta)|^2\rangle^{1/2}}
\end{equation} 
where $\mathds{P}$ is the Leray divergence-free projector given by $\mathds{P}(\vec{v}) = \vec{v} - \nabla \Delta^{-1}(\nabla \cdot \vec{v})$ and $\langle \cdot \rangle$ is the spatial average. These flows will be studied numerically later and is the main focus of this paper.

We introduce the following measures as useful observables of mixing over time. we use the $H^{-1}$ norm to define the (exponential) rate of mixing as
\begin{equation}
\label{eq:rate}
r(t) = -  \frac{\ddt{}\hmone{\theta}}{\hmone{\theta}}.
\end{equation}
We define the following ratio as a measure of the characteristic filamentation length scale:
\begin{equation}
\lambda(t)\equiv  2\pi \frac{\|\nabla^{-1}\theta(\,\cdot\,,t)\|_{L^{2}}}{\|\theta(\,\cdot\,,t)\|_{L^{2}}}.
\end{equation}
Note that if the tracer concentration field is composed of only one Fourier mode with wave number $\vec{k}$ (i.e. $\theta(\vec{x},t) = Re[ A e^{-i\vec{k}\cdot \vec{x}}]$ where $A$ is a complex constant), then $\lambda(t)$ returns the wavelength of the wave number $\vec{k}$. In general, $\lambda$ is the weighted root mean square wavelength with weights given by $|\theta_{\vec{k}}|/\ltwo{\theta}$.

\subsection{Shell model predictions of local-in-time optimisation}

The shell model is a model that mimics the spectral dynamics present in the advection-diffusion equation. The model consists of a system of ordinary differential equations with nearest-neighbour coupling between `shells' in wave number space. \cite{Miles2017a} performed local-in-time mixing optimisation in this model. The shell-model analysis predicts a limiting length scale given by the Batchelor scale, $\Lambda_{\Gamma} =\sqrt{\frac{\kappa}{\Gamma}}$  and its generalisation $\Lambda_{U}= \frac{U}{\kappa} $. The non-dimensional versions are given by $\lambda_{\Gamma}= \frac{1}{\sqrt{Pe}}$ and $\lambda_{U} = \frac{1}{Pe}$. From here forward, we will refer to the Batchelor scale to mean either $\lambda_{\Gamma}$ or its generalisation $\lambda_{U}$.  The predicted long-term rates (after reaching the Batchelor scale) are given by $R_{\Gamma} =\kappa/\lambda_{\Gamma}^2 $  and  $R_{U}=\kappa/\lambda_{U}^2$.  The non-dimensional versions are given by $r_{\Gamma} =1$ and $r_{U} = Pe $.

\subsection{Bounds for $L^{\infty}$ constrained flows}
We consider a subset of $L^{2}$ constrained flows --- those belonging to $L^{\infty}$. In this restricted setting the rate-of-strain and speed are bounded point-wise uniformly in space and time rather than demanding that they merely be $L^2$ integrable as before. We will provide bounds on $\lambda$ and measures of mixing in this restricted setting. 

\label{sec:linfty_flows}
\subsubsection{Results for $\linf{\nabla \vec{u}} = 1$}

The time derivative of $\lambda^2$ is
\begin{equation*}
	\ddt{\lambda^2} = \frac{2}{Pe}
		\left[ 
			\frac{\hone{\theta}^2\hmone{\theta}^2}
					{\ltwo{\theta}^4}  
			- 1
		\right]
		+ 2 \frac{\sint{\nabla^{-1}\theta \cdot \nabla\vec{u} \cdot 
							\nabla^{-1}\theta  }}
					  {\ltwo{\theta}^{2}}
\end{equation*}
and by H\"older's inequality, we deduce
\begin{equation*}
\label{eq:length_ineq_rate-of-strain}
	\ddt{\lambda^2} \geq \frac{2}{Pe} \left[ 
			\frac{\hone{\theta}^2\hmone{\theta}^2}
					{\ltwo{\theta}^4}  
			- 1
		\right] - 2  \lambda^2 .
\end{equation*}
This establishes a lower bound on $\lambda$ at each instant: by apply Gr\"onwall's inequality and the fact that the bracketed term is greater than or equal to zero, it follows that
\begin{equation}
\label{eq:exponential_enstrophy}
	\lambda (t) \geq \lambda(0)e^{- t}.
\end{equation}
Therefore, perfect mixing in finite time is impossible for bounded rate-of-strain flows.

Furthermore,
 \begin{eqnarray*}
\frac{d}{dt}\left(\frac{\|\nabla\theta\|_{L^{2}}^2}{\|\theta\|_{L^{2}}^2}\right) &= \frac{\|\theta\|_{L^{2}}^2\frac{d}{dt}\|\nabla\theta\|_{L^{2}}^2-\|\nabla\theta\|_{L^{2}}^2\frac{d}{dt}\|\theta\|_{L^{2}}^2}{\|\theta\|_{L^{2}}^4}\\
&= \frac{-2\int \partial_{i}u_{j}\partial_{i}\theta\partial_{j}\theta - \frac{2}{Pe} \|\Delta\theta\|_{L^{2}}^2}{\|\theta\|_{L^{2}}^2}+\frac{2}{Pe}\frac{\|\nabla\theta\|_{L^{2}}^4}{\|\theta\|_{L^{2}}^4} \\
&=-\frac{2}{Pe}\left(\frac{\|\Delta\theta\|_{L^{2}}^2}{\|\theta\|_{L^{2}}^2} - \frac{\|\nabla\theta\|_{L^{2}}^4}{\|\theta\|_{L^{2}}^4} \right) - 2\frac{\sint{\nabla\theta \cdot \nabla\vec{u} \cdot \nabla\theta  }}{\|\theta\|_{L^{2}}^2} 
\\
&\leq 2 \frac{\hone{\theta}^2}{\ltwo{\theta}^2}
\end{eqnarray*}
and using $\ddt{}\ltwo{\theta}^2 = -\frac{2}{Pe} \hone{\theta}^2$, it follows that
\begin{equation}
\ltwo{\theta}\geq  \ltwo{\theta_{0}}\exp\left[-\frac{1}{2Pe}\frac{\hone{\theta_{0}}^2}{\ltwo{\theta_{0}}^2}\left(e^{2 t} -1\right)\right].
\end{equation}
Using this with \eref{eq:exponential_enstrophy}, we deduce the lower bound
\begin{equation}
\hmone{\theta} \geq  \hmone{\theta_{0}} \exp\left[- t -\frac{1}{2 Pe}\frac{\hone{\theta_{0}}^2}{\ltwo{\theta_{0}}}\left(e^{2 t} -1\right)\right].
\end{equation}

\subsubsection{Results for $\linf{\u}= 1$}
Here we extend the result of \cite{Chi-Cheu1996} to show that the presence of diffusion rules out perfect mixing in finite time for bounded velocity flows.  First note that
\begin{eqnarray*}
	 \hone{\theta}^2 &=& - 2\sint{\theta \lap \theta} \\
	 							&=& Pe \sint{\theta\left(\ppt{\theta}
	 									-\frac{1}{Pe}\lap \theta\right)} 
	 									-Pe \sint{\theta\left(\ppt{\theta}
	 									+\frac{1}{Pe}\lap \theta\right)},
\end{eqnarray*}
\begin{eqnarray*}
	\ddt{}\ltwo{\theta}^2 &=& 2\sint{\theta\ppt{\theta}} \\
										 &=&\sint{\theta\left(\ppt{\theta}
	 									-\frac{1}{Pe}\lap \theta\right)} 
										 + \sint{\theta\left(\ppt{\theta}
	 									+\frac{1}{Pe}\lap \theta\right)} ,
\end{eqnarray*}
and
\begin{eqnarray*}
	\ddt{}\hone{\theta}^2 &=& -2\sint{\ppt{\theta}\lap \theta} \\
	 									&=& Pe \sint{\left(\ppt{\theta}
	 									-\frac{1}{Pe}\lap \theta\right)^2} 
	 									-Pe \sint{\left(\ppt{\theta}
	 									+\frac{1}{Pe}\lap \theta\right)^2} .
\end{eqnarray*}

Then compute:
\begin{eqnarray*}
	\ddt{} \pbrac{ \frac{\hone{\theta}^2}{\ltwo{\theta}^2} } 
			&=& \frac{1}{\ltwo{\theta}^4}
			\sbrac{
				\ltwo{\theta}^2\ddt{}\hone{\theta}^2
				-\ddt{}\ltwo{\theta}^2\hone{\theta}^2			
			}\\
			&=& \frac{1}{\ltwo{\theta}^2}
			 \Bigg[ Pe \sint{\left(\ppt{\theta} -\frac{1}{Pe}\lap \theta\right)^2}  \\
 				& & \qquad\qquad\qquad \qquad \qquad 
				-Pe\sint{\left(\ppt{\theta}+\frac{1}{Pe}\lap \theta\right)^2} 
			\Bigg]\\
		        &-&\frac{1}{\ltwo{\theta}^4}
			\Bigg[ 
				Pe \pbrac{\sint{\theta\left(\ppt{\theta}-\frac{1}{Pe}\lap \theta\right)} }^2\\
 				& & \qquad\qquad\qquad \qquad
				-Pe\pbrac{
					 \sint{\theta\left(\ppt{\theta}
	 									+\frac{1}{Pe}\lap \theta\right)} 
 				}^2					
			\Bigg].
\end{eqnarray*}
Using H\"older's inequality and \eref{eq:PDE_advection}, this simplifies to
\begin{eqnarray*}
	\ddt{} \pbrac{ \frac{\hone{\theta}^2}{\ltwo{\theta}^2} } 
			&\leq & \frac{Pe}{\ltwo{\theta}^2}
			\sbrac{
					 \sint{(\vec{u}\cdot \nabla \theta)^2} 
			}.
\end{eqnarray*}
Again applying H\"older's inequality we have
\begin{equation}
\label{eq:k2growth_energy}
	\ddt{} 
		\pbrac{ 
			\frac{\hone{\theta}^2}{\ltwo{\theta}^2} 
		} 
		\leq  
		Pe
		\frac{\hone{\theta}^2}{\ltwo{\theta}^2} 
\end{equation}
and thus 
\begin{equation*}
		\frac{\hone{\theta}}{\ltwo{\theta}} 
		\leq  
		\frac{\hone{\theta_0}}{\ltwo{\theta_0}}
		\exp{\pbrac{\frac{Pe}{2} t}}.
\end{equation*}
The inequality $\hone{\theta}\hmone{\theta}\geq \ltwo{\theta}^2$ then ensures that
\begin{equation}
\label{eq:lambda_bound}
\lambda(t) \geq \frac{\ltwo{\theta_0}}{\hone{\theta_0}}\exp{\pbrac{-\frac{Pe}{2}t}}.
\end{equation}
Using \eref{eq:k2growth_energy} together with  $\ddt{}\ltwo{\theta}^2 = -\frac{2}{Pe} \hone{\theta}^2$ we observe that
\begin{equation*}
\ltwo{\theta}\geq \ltwo{\theta_{0}}\exp\left[-\frac{1}{Pe^2}\frac{\hone{\theta_0}^2}{\ltwo{\theta_0}^2}\left(e^{Pe \, \, t}-1\right)\right]
\end{equation*}
and this combined with  \eref{eq:lambda_bound} implies
\begin{equation}
\hmone{\theta}\geq \frac{\ltwo{\theta_{0}}^2}{\hone{\theta_0}}\exp\left[-\frac{Pe}{2} \,\, t-\frac{1}{Pe^2}\frac{\hone{\theta_0}^2}{\ltwo{\theta_0}^2}\left(e^{Pe \,\, t}-1\right)\right].
\end{equation}

\section{Numerical experiment: local-in-time optimisation}
\label{sec:numerical_experiment}
\subsection{Methodology}

We solve \eref{eq:nd_ade} by using a Fourier basis to represent the discretised spatial domain with a 4th order Runge-Kutta time-stepping method. All simulation code was created in the programming language Python with package modules, pyfftw and numpy. The code repository can be found at \href{http://github.com/cjm715/lit}{ http://github.com/cjm715/lit}.


\subsection{Results}

\begin{figure}
\includegraphics[width=\textwidth]{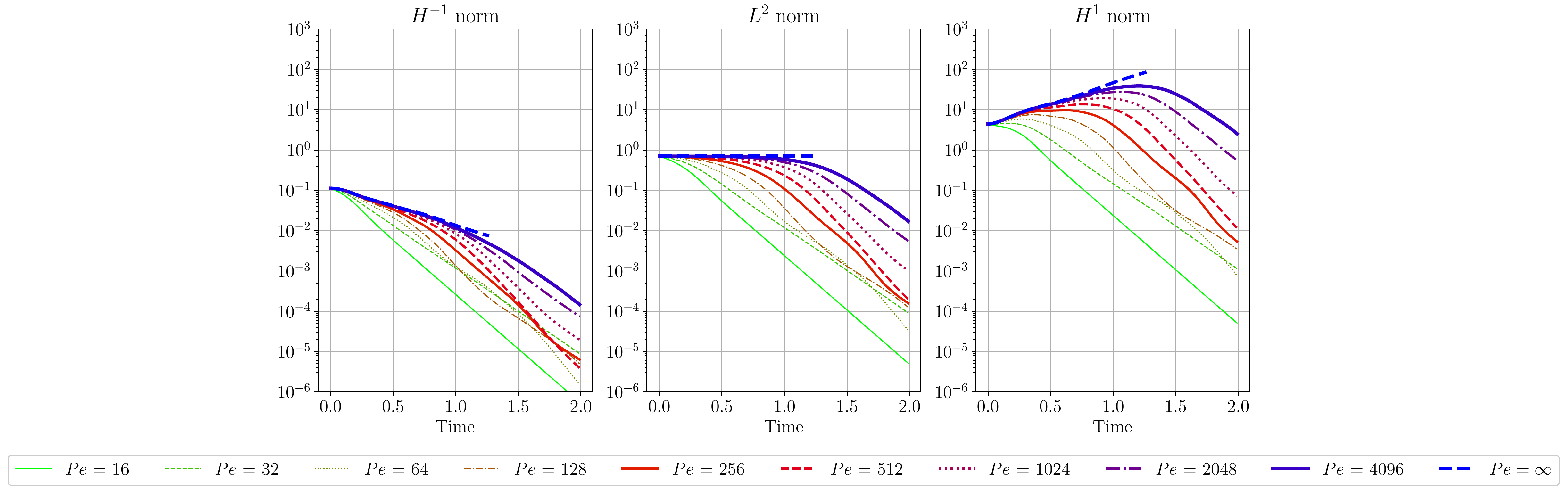}
\caption{$H^{-1}, L^{2},$ and $H^{1}$ norms of the concentration field under the optimal enstrophy-constrained flow.  }
\label{fig:enstrophy_norms}
\end{figure}
\begin{figure}
\includegraphics[width=\textwidth]{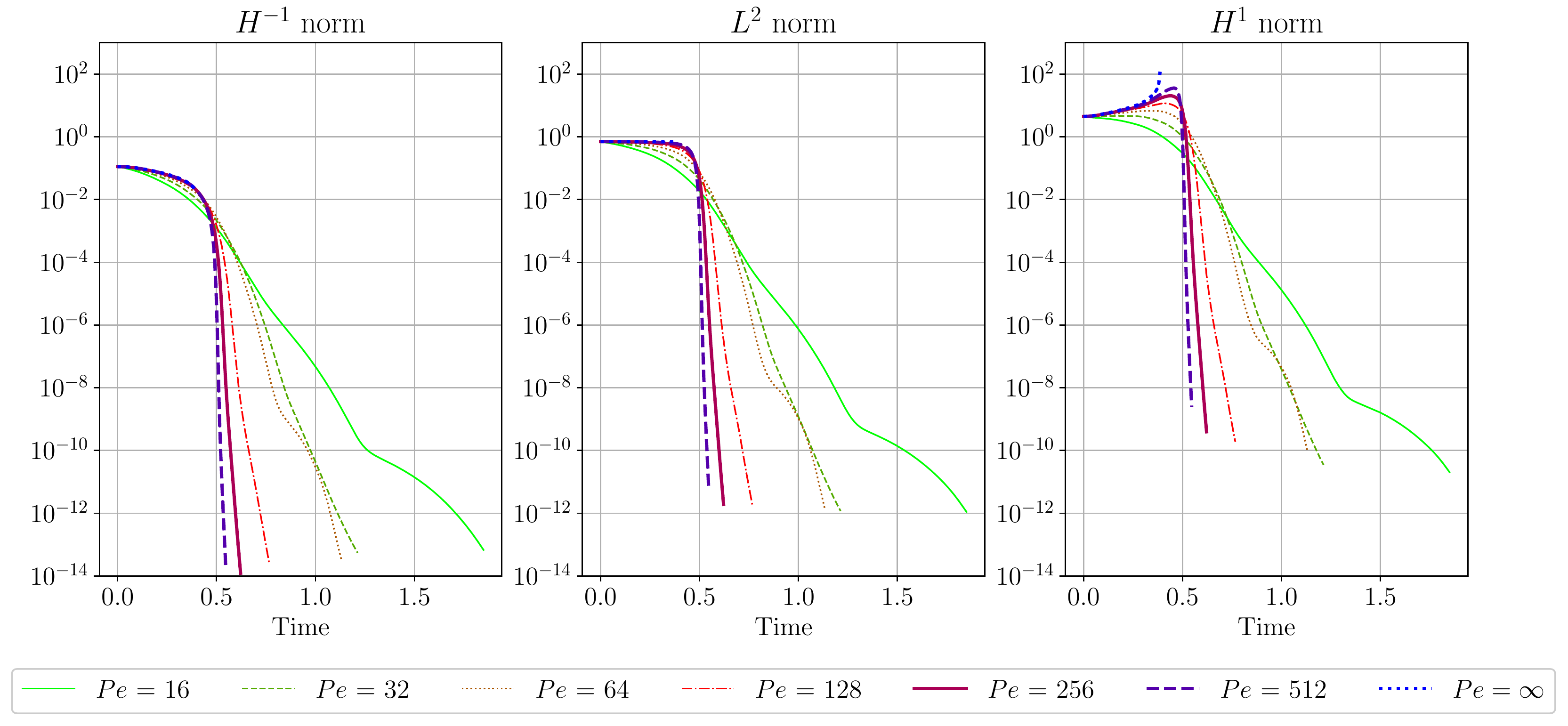}
\caption{$H^{-1}, L^{2},$ and $H^{1}$ norms of the concentration field under the optimal energy-constrained flow.}
\label{fig:energy_norms}
\end{figure}

We now investigate the mixing performance under the local-in-time optimal flows. Figure \ref{fig:enstrophy_norms} and \ref{fig:energy_norms} show how the different mixing measures ($H^{-1}$, $L^2$, and $H^{1}$ norms) vary in time for different values of $Pe$ for the enstrophy and energy constrained cases respectively. Notice how the long-term mixing rate appears to be exponential for all three mixing measures. This exponential rate is consistent with shell model predictions, yet weaker than the double-exponential decay rate derived by the $L^{\infty}$ constrained flow analysis.

\begin{figure}
\includegraphics[width=\textwidth]{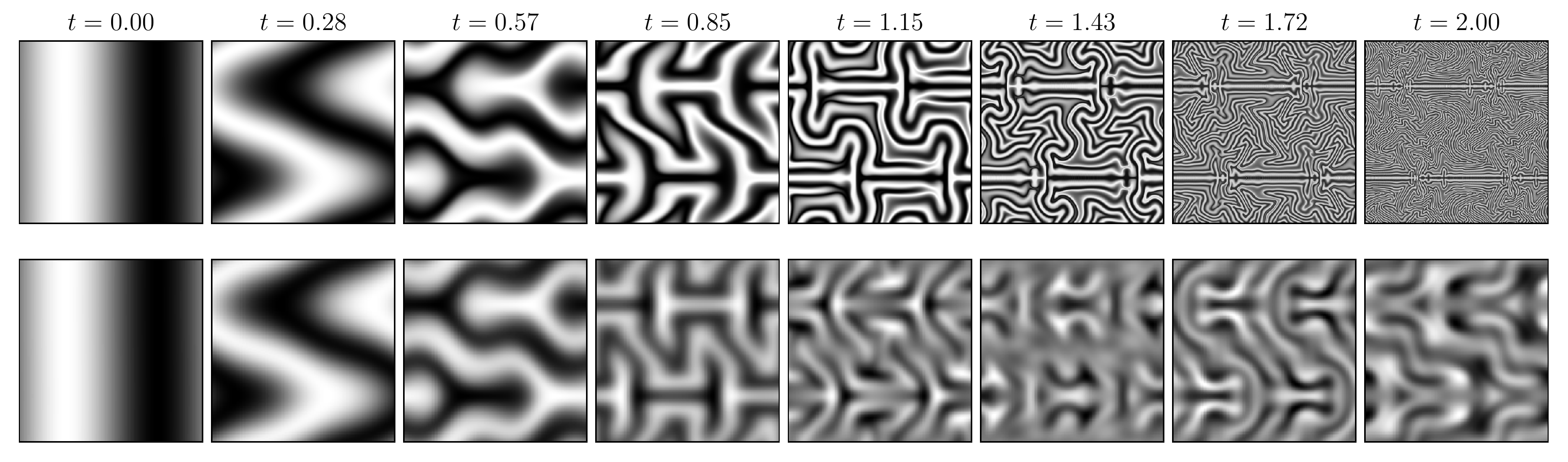}
\caption{Local-in-time optimisation with enstrophy constraint. Top filmstrip is for $Pe =\infty$ and the bottom filmstrip is $Pe=256$. Note that the grey-scale for the $Pe=\infty$ is constant in time while it is adjusted to show the tracer concentration structure in the finite $Pe$ case. }
\label{fig:enstrophy_film}
\end{figure}
\begin{figure}
\includegraphics[width=\textwidth]{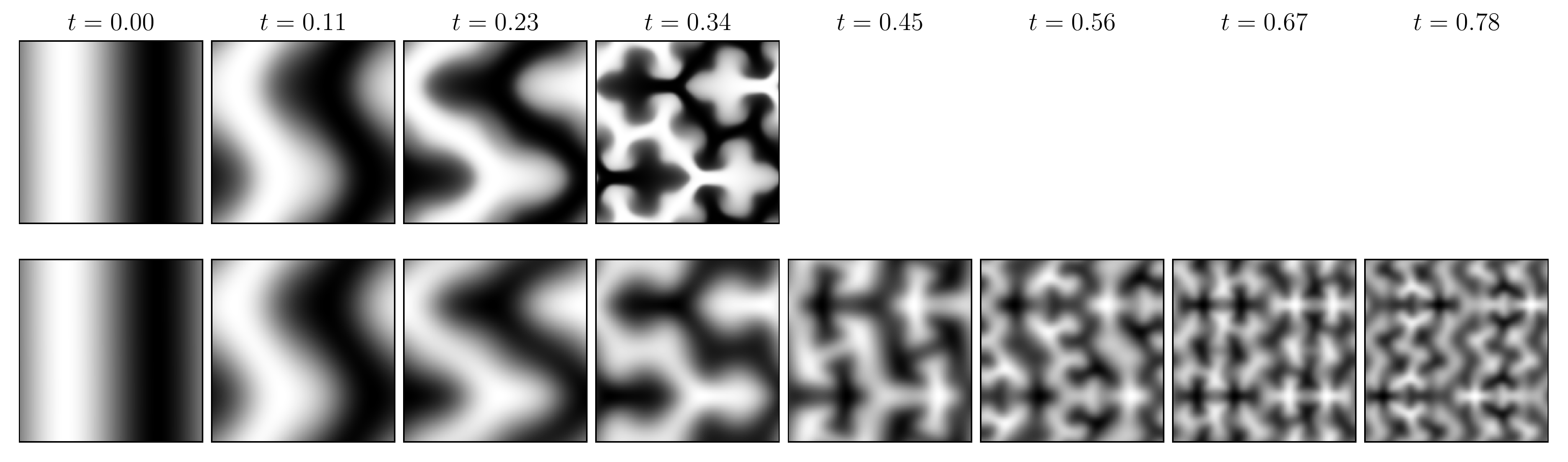}
\caption{Local-in-time optimisation with energy constraint. Top filmstrip is for $Pe = \infty$ and the bottom filmstrip is $Pe=32$. Note that the grey-scale for the $Pe=\infty$ is constant in time while it is adjusted to show the tracer concentration structure in the finite $Pe$ case. The numerical computation is truncated at time $t=0.34$ due to length scales rapidly decreasing past the grid size resolution immediately after $t=0.34$. Fixed energy constrained flows that produce infinitesimally small lengths in finite time have been constructed \cite{JMP2012}. We suspect that the same phenomena may be occurring here.}
\label{fig:energy_film}
\end{figure}
\begin{figure}
\includegraphics[width=\textwidth]{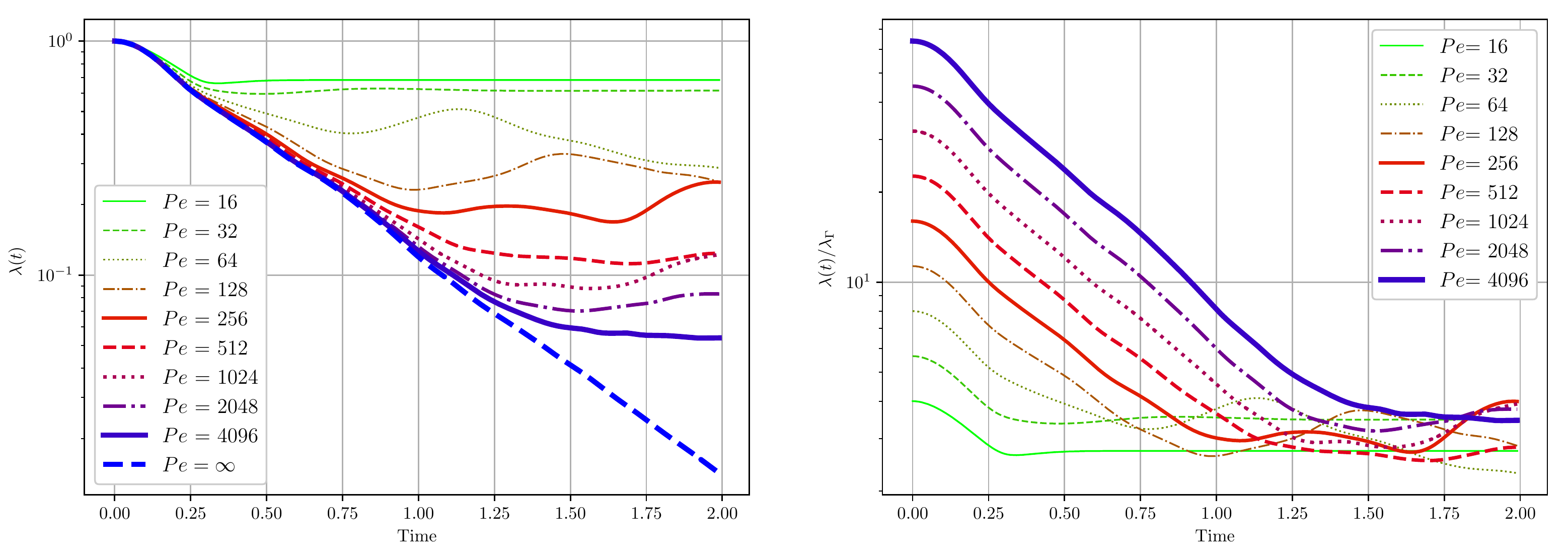}
\caption{The left subplot shows the filament length $\lambda$ over time subject to the optimal enstrophy-constrained flow. The right subplot is the same data except scaled: $\lambda(t)/\lambda_{\Gamma} = \lambda(t)\sqrt{Pe}$.}
\label{fig:enstrophy_length}
\end{figure}
\begin{figure}
\includegraphics[width=\textwidth]{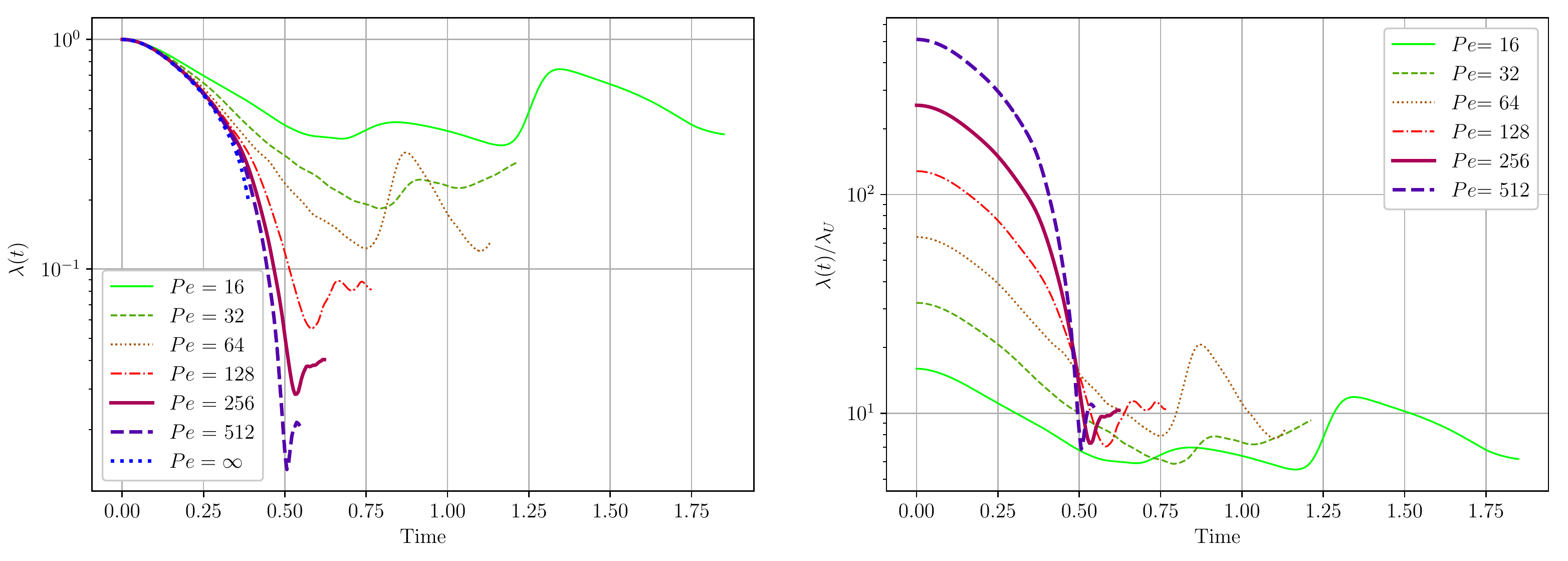}
\caption{The left subplot shows the filament length $\lambda$ over time subject to the optimal energy-constrained flow. The right subplot is the same data except scaled: $\lambda(t)/\lambda_{U} = \lambda(t) Pe$.}
\label{fig:energy_length}
\end{figure}
\begin{figure}
\centering
\includegraphics[width=0.5\textwidth]{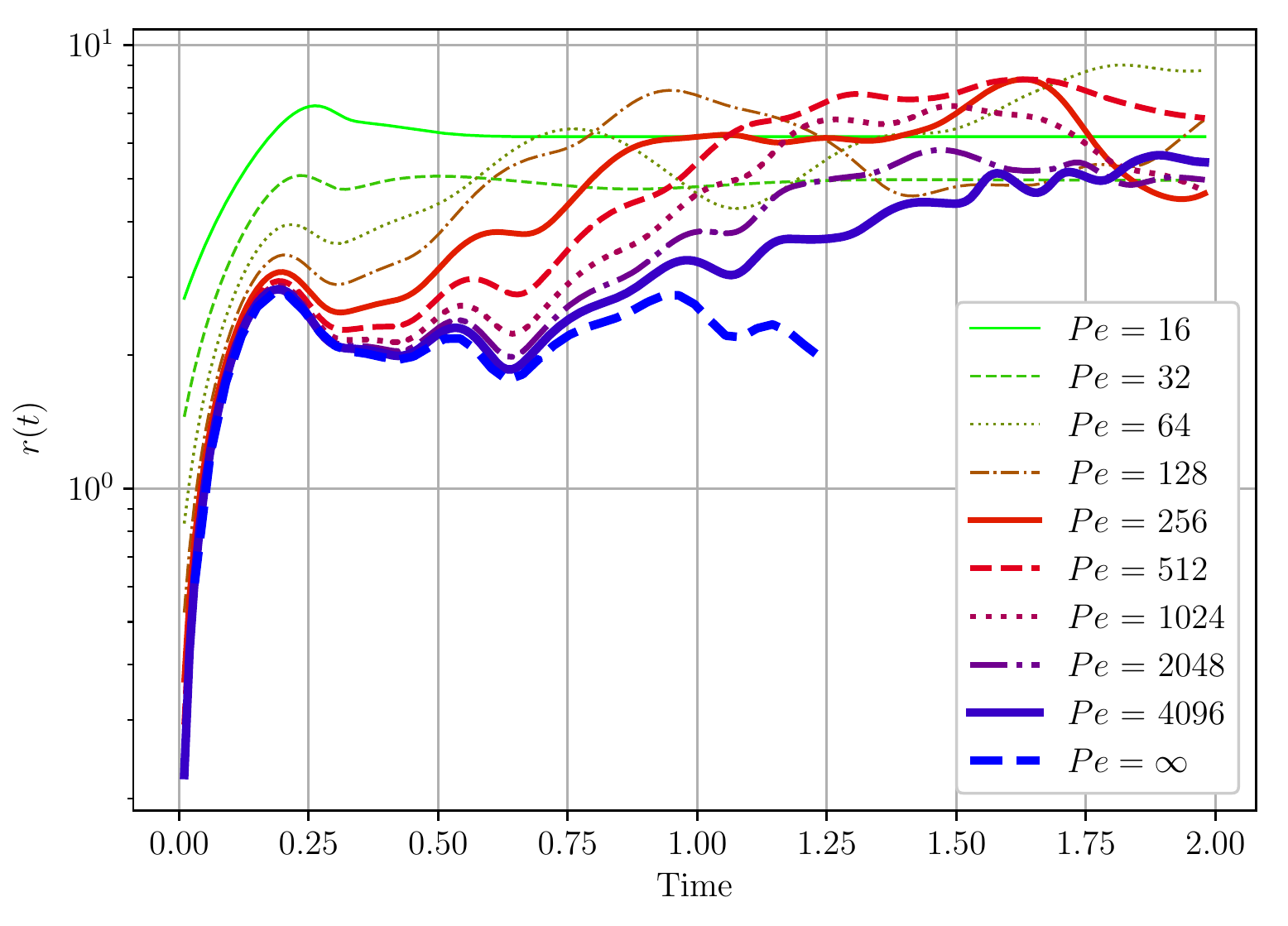}
\caption{Mixing rate $r(t)$ over time when subject to the optimal enstrophy-constrained flow.}
\label{fig:enstrophy_rate}
\end{figure}
\begin{figure}
\centering
\includegraphics[width=\textwidth]{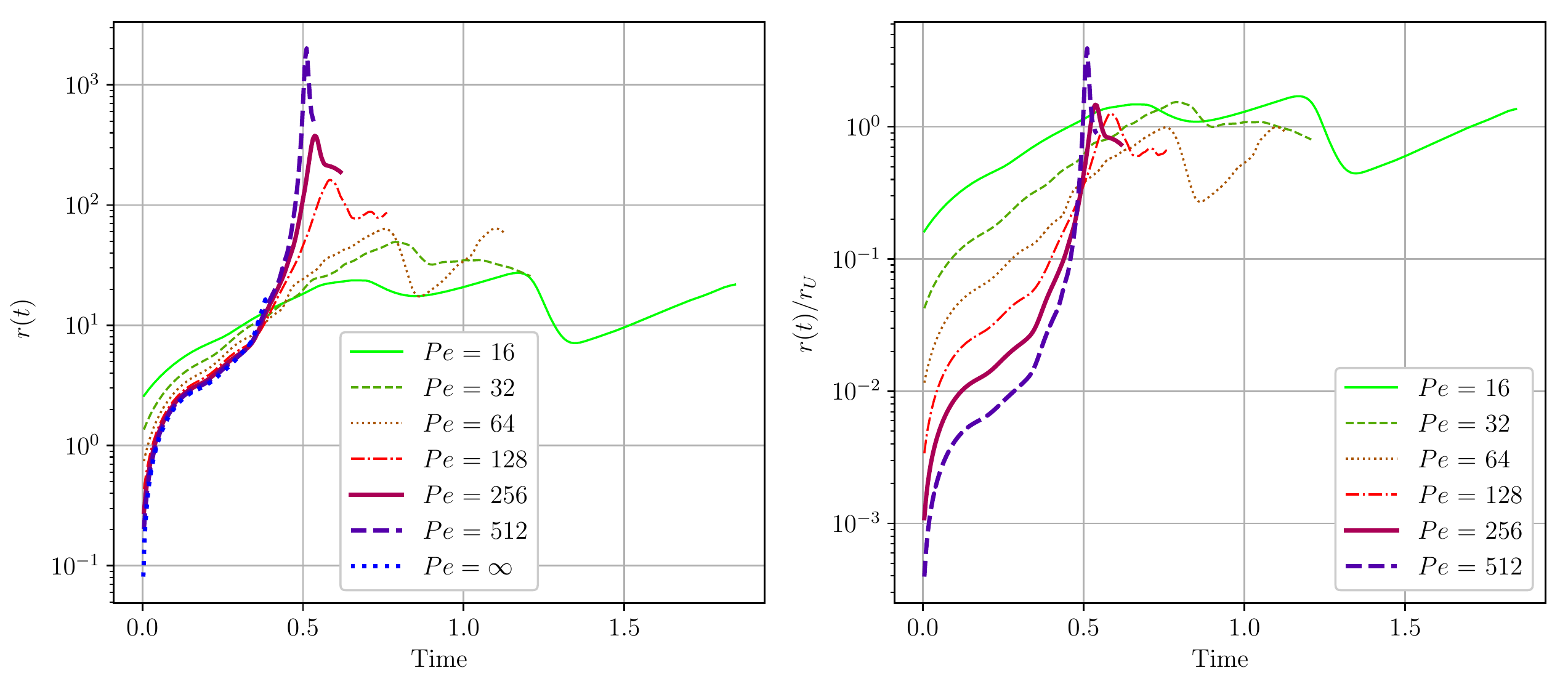}
\caption{The left subplot shows the mixing rate $r(t)$ over time when subject to the optimal energy-constrained flow. The right subplot is the same data except scaled: $r(t)/r_{U} = r(t)/Pe$.}
\label{fig:energy_rate}
\end{figure}

 Figure \ref{fig:enstrophy_film} shows the evolution of a scalar field under the optimal flow for the enstrophy constraint. The top film strip corresponds to $Pe =\infty$ while the bottom is $Pe = 256$. The time evolution is initially similar but soon diverges over time. Figure \ref{fig:energy_film} shows the evolution for the energy constraint. The top film strip corresponds to $Pe =\infty$ while the bottom is $Pe = 32$. Notice that, unlike the $Pe = \infty$ cases, the flows with finite $Pe$ are incapable of creating length scales arbitrarily small for either the energy or enstrophy cases.  The left subplot of Figures \ref{fig:enstrophy_length} and \ref{fig:energy_length} shows this phenomena more quantitatively by showing $\lambda$ over time eventually reaching a plateau. The shell-model prediction of this limiting length scale is the Batchelor scale given by $\lambda_{\Gamma} = 1/\sqrt{Pe}$ for the enstrophy case and  $\lambda_{U} = 1/Pe$ for the energy case. The right plots of Figures \ref{fig:enstrophy_length} and \ref{fig:energy_length} shows scaled versions of $\lambda$ given by  $\lambda/\lambda_{\Gamma}$ and $\lambda/\lambda_{U}$ respectively.  Notice how they plateau around an $O(1)$ constant. Thus this result is consistent with the shell-model predictions. 
   
The mixing rates for the enstrophy case are shown in Figure \ref{fig:enstrophy_rate}. The rate during the transient phase is $\Gamma$ which is consistent with rates expected from $Pe=\infty$ mixing studies. For all $Pe$ considered, there is an increase in the rate of mixing after transient behaviour has finished to a long-term rate. Perhaps surprisingly, this long-term mixing rate appears to be {\it independent} of $Pe$ for fixed enstrophy. This suggests that the optimal long-term rate of mixing is only dependent on the rate-of-strain $\Gamma$ and not influenced by the strength of diffusion. 

It should be noted that the onset of the long-term rate is affected by the value of $Pe$. When there is strong diffusion (small $Pe$), the Batchelor scale is reached quickly. From the work of G. Iyer {\it et al.} \cite{GI2014} and C. Seis \cite{CS2013}, we know that $\lambda$ decreases at most exponentially for $Pe = \infty$. If we assume that the local-in-time optimal flows nearly saturate this bound in the transient phase, we model $\lambda$ as  $\lambda (t) = \lambda(0)\exp(- \alpha t) $ during this time. We expect the critical transition time $t_{c}$ that marks the end of this transient period to satisfy $\lambda(t_{c})= \lambda_{\Gamma}$. This time is theorised to be $t_{c}=\frac{1}{\alpha}\ln(\lambda(0)/\lambda_{\Gamma}) = \frac{1}{\alpha}\ln ( \sqrt{Pe} )$ for $Pe>1$ (If $Pe \leq 1$, then there is no transient phase). Hence, a smaller value of $Pe$ will result in an earlier onset of the long-term rate of mixing. Therefore, it is advantageous to have strong diffusion (small $Pe$) so that there is an earlier onset of the long-term mixing rate (although independent of $Pe$) which is an improvement over the mixing rate of the purely non-diffusive situation ($Pe=\infty$).
 
For the energy case, the long-term mixing rate decreases with decreasing $Pe$ (see the left subplot of figure \ref{fig:energy_rate}).  {\it Thus, strong diffusion results in a weak long-term mixing rate}. The right subplot of Figure \ref{fig:energy_rate} is $r/r_{U} =  r/Pe$. We see oscillations of $r/r_{U}$ around a value that is $O(1)$ which indicates that our numerical results are consistent with our predictions from the shell model. Thus, the long-term mixing rate is proportional to $Pe$ in contrast to the long-term mixing rate of enstrophy which carries no dependence on $Pe$.

For the energy case, the onset of the long run-mixing behaviour can be determined by the following model. From the work of E. Lunasin {\it et al.} \cite{JMP2012} on the fixed energy case, $\lambda(t)$ can decrease linearly in time to produce perfect mixing in finite time. We model the transient phase as $\lambda(t)=\lambda(0)(1-\beta t)$. Therefore, we theorise that the critical transition time is $t_{c}=\frac{1}{\beta}(1 -\lambda_{U}/\lambda(0)) = \frac{1}{\beta}(1 - 1/Pe)$ with $Pe> 1$ (If $Pe \leq 1$, there is no transient phase) for the energy case. Thus, it is true that on can still achieve an earlier onset of the long-term mixing behaviour by choosing a smaller $Pe$. However, an earlier onset time is accompanied by a slower long-term mixing rate. As for choosing a large $Pe$, the onset time is bounded above by $\frac{1}{\beta}$ and results in a faster long-term mixing rate. Thus, it is advantageous to have weak diffusion (large $Pe$) for mixing in the fixed energy case. This benefit is well illustrated by $H^{-1}$ norm in figure \ref{fig:energy_norms}. Notice that the mixing rate is initially slow for $Pe = 512$ but then out competes the mixing rate of smaller values of $Pe$.    


\section{Discussion}
\label{sec:discussion}
The local-in-time optimisation results suggest that there is a limiting length scale for passive tracer mixing whenever $L^{2}$ flows are instantaneously optimised to decrease the mix-norm. The bounds derived under the $L^{\infty}$ constrained flow assumption did not result in proving this observation, but they did definitively rule out the possibility of perfect mixing in finite time for $L^{\infty}$ flow constraints. 

We suspect that the bounds obtained for $L^{\infty}$ flows are not sharp and could be improved further. The $L^{\infty}$ flow analysis produced a double-exponential lower bound on the $H^{-1}$ mix-norm rather than exponential as possibly expected given the numerical results for local-in-time optimal $L^2$ flows. The double-exponential bounds arise from the use of exponential upper bounds on the quantity $\frac{\hone{\theta}}{\ltwo{\theta}}$ in time for both $L^{\infty}$ flow constraints considered. We surmise that $\frac{\hone{\theta}}{\ltwo{\theta}} < C$ (where $C$ is a constant) for all time $t$ as supported by our numerical results and may be proven by exploiting the incompressiblity condition. If this is true generally for all $L^{\infty}$  flows, then our previous analysis would demonstrate that the $H^{-1}$ mix-norm is bounded below by an exponential.

Note that the pure diffusive case discussed in the introduction can always be employed as a mixing strategy by simply not having a flow field at all ($\vec{u} =\vec{0}$) provided that the flow intensity constraints are generalised to inequalities such as $\|\vec{u}\| \leq UL^{d/2}$ and $\|\nabla \vec{u}\| \leq \Gamma L^{d/2}$.This is a valuable strategy if one is content with mixing at a long-term rate of $\kappa k_{\min}^2$ where $k_{\min} = \min \{ |\vec{k}|  :  |\hat{\theta}_{\vec{k}}(0)| > 0  \}$. This may be advised in fact if $k_{\min} > 2 \pi / \lambda_{B}$. This may well be the most optimal strategy. Invoking a flow may cause the lower wave number modes to become `populated' and therefore may limit the mixing rate. It is important to keep this simple strategy in mind when trying to rigorously prove bounds on the mix-norm. This strategy has an important implication --- there does not exist a lower bound on the mix-norm of the form $\hmone{\theta} \geq A e^{-rt}$ where  $r$ is independent of the initial data.

In future work, we would like to consider the optimal control problem with finite-time optimisation to minimise the $H^{-1}$ mix-norm at the end time rather than instantaneously attempting to minimise its decay rate. This might lead to flows that can produce even smaller length scales. In Miles and Doering \cite{Miles2017a}, finite-time optimisation was explored in the context of the shell model where it was found that global-in-time and local-in-time optimisation appeared to give similar mixing rates. For the shell model, however, the analysis was consistent with computation. In the partial differential equation case, the gap between analysis and computation remains to be closed.


\section{Conclusion}
\label{sec:conclusion}

Our numerical study of local-in-time optimisation suggests that there is a limiting length scale, a generalised Batchelor length scale, which in turn determines a long-term mixing ``Batchelor rate". In dimensional form, this Batchelor rate was found to be proportional to $\Gamma$ for the fixed enstrophy case and $U^{2}/\kappa$ for the fixed energy case. These rates are consistent with those found in the context of the shell model. Although the Batchelor scale has been a theorised lower bound on the length scales present on turbulent flows, it has not been proven rigorously. We hope this numerical study provides insight and promotes investigation into mathematically proving what conditions are necessary on the flow for a length scale limitation. This is especially important since it plays a crucial role in the achievable mixing rates. Furthermore, we provided numerical evidence that (1), for fixed enstrophy optimal flows, strong diffusion can benefit from an early onset of a long-term mixing rate (where the rate itself however is independent of diffusion strength) while (2), for energy fixed optimal flows, strong diffusion weakens the long-term mixing rate.

\section*{Acknowledgements}
We gratefully acknowledge helpful comments from Hongjie Dong, Luis Escauriaza, Guatum Iyer, Alexander Kiselev, Anna Mazzucato, Christian Seis, Ian Tobasco, and Karen Zaya. This work was supported in part by NSF Awards PHY-1205219 and DMS-1515161. One of us (CRD) is additionally grateful for Fellowship funding from the John Simon Guggenheim Memorial Foundation.

\section*{References}
\bibliographystyle{iopart-num}
\bibliography{library}

\end{document}